\documentclass[traditabstract]{aa}
\usepackage{epsfig}
\input psfig.sty
\begin{document}

\title{Constraints on the duality relation from ACT cluster data}
\author{
R. \ S.\ Gon\c calves \inst{1}\thanks{\email{r.de-sousa-goncalves@imperial.ac.uk}}
\and
A.\ Bernui. \inst{2}\thanks{\email{bernui@on.br}}
\and
R. \ F. \ L.\ Holanda. \inst{2,3}\thanks{\email{holanda@uepb.edu.br}}
\and
J.\ S.\ Alcaniz \inst{2}\thanks{\email{alcaniz@on.br}}
}
\institute{$^1$Department of Physics, Imperial College, Blackett Laboratory, London SW7 2AZ, United Kingdon\\ $^{2}$Departamento de Astronomia, Observat\'orio Nacional, 20921-400, Rio de Janeiro - RJ, Brasil \\ $^{3}$Departamento de F\'{\i}sica, Universidade Estadual da Para\'{\i}ba, 58429-500, Campina Grande - PB, Brasil and $^{4}$Departamento de F\'{\i}sica, Universidade Federal de Campina Grande, 58429-900, Campina Grande - PB, Brasil }
         

\date{}

\abstract {The cosmic distance-duality relation (CDDR), $d_L(z) (1 + z)^{2}/d_{A}(z) = \eta$, where $\eta = 1$ and $d_L(z)$ and $d_A(z)$ are, respectively, the luminosity and
the angular diameter distances, holds as long as the number of photons is conserved and gravity is described by a metric theory. Testing such hypotheses is, therefore, an important task for both cosmology and fundamental physics. In this paper we use 91 measurements of the gas mass fraction of galaxy clusters recently reported by the Atacama Cosmology  Telescope (ACT) survey along with type Ia supernovae observations of the Union2.1 compilation to probe a possible deviation from the value $\eta = 1$. Although in agreement with the standard hyphothesis, we find that this combination of data tends to favor negative values of $\eta$ which might be associated with some physical processes increasing the number of photons and modifying the above relation to $d_L < (1+z)^2d_A$.}


\maketitle


\section{Introduction}

The so-called distance reciprocity law, proved long ago by Etherington (Etherington, 1933) (see also Ellis, 1971, 2007), is a fundamental keystone for the interpretation of observational data in cosmology. Considering that the number of photons from a given source is conserved, it provides the following relation between the angular ($d_{\scriptstyle A}$) and luminosity ($d_{\scriptstyle L}$) distances:
\begin{equation}
  \frac{D_{\scriptstyle L}}{D_{\scriptstyle A}}{(1+z)}^{-2}= \eta \quad \mbox{with} \quad \eta = 1\; .
  \label{rec}
\end{equation}
This version of the reciprocity relation, also known as the cosmic distance-duality relation (CDDR), is valid for {all} cosmological models based on Riemannian geometry, only requiring that source and observer are connected by null geodesics in a Riemannian spacetime and that the number of photons be conserved. Examples of non-standard frameworks that violate the CDDR include scenarios where photons do not travel on unique null geodesics (Csaki {\it{et al.}} 2002), models with variations of fundamental constants (Brax {\it{et al.}} 2013), with photon non-conservation due to coupling to particles beyond the standard model of particle physics (Avgoustidis {\it{et al.}}2010, 2012), absorption by dust (Basset \& Kunz 2004), among others (see Uzan {\it{et al.}} 2004 and references therein).

Recently, several authors have explored different techniques to test the CDDR. For instance,   Bassett \& Kunz (2004) used type Ia supernovae (SNe Ia) data as measurements of $d_L$ and  estimates of angular distances from FRIIb radio galaxies (Daly \& Djorgovski 2003) and ultra-compact radio 
sources (Gurvitz 1994; 1999; Lima \& Alcaniz 2002) to test possible deviations of the CDDR. Perhaps due to lensing  magnification bias, they found a 2$\sigma$ violation caused by an excess in brightening of SNIa at $z > 0.5$. Ellis {\it{et al.}}  (2013) proposed an interesting test to Eq. (\ref{rec}) using the CMB spectrum. From this observable, it was found that the CDDR relation cannot  be violated by more than 0.01\% between the decoupling era and today. Khedekar \& Chakraborti (2011) proposed the use of a redshifted 21 cm signal from disk galaxies, where neutral hydrogen (HI) masses are seen to be almost linearly correlated with surface area,  to detect a possible violation of the CDDR.  

Measurements of the angular diameter distance from galaxy clusters observations, calculated through their X-ray and Sunyaev-Zeldovich observations, have also been widely used to test the CDDR (see. e.g., Holanda {\it{et al.}} (2012a) and references therein). It is known, however, that the verification of the CDDR validity depends on the assumptions used to describe the galaxy clusters (Holanda, Lima \& Ribeiro, 2010,2012; Nair, Jhingan \& Jain 2011; Li, Wu \& Yu 2011; Meng, Zhang \& Zhan 2012; Lima, Cunha \& Zanchin 2012; Xi {\it{et al.}} 2013).  As an example, assuming the CDDR, Holanda {\it{et al.}} (2012b) and Meng {\it{et al.}} (2012) showed that the elliptical $\beta$-model constitutes a better geometrical description of the galaxy cluster structure when compared to the spherical $\beta$-model. Following a different approach, Gon\c calves {\it{et al.}} (2012) showed that measurements of  gas mass fraction ($f_{gas}$) of galaxy clusters depend on the validity of the CDDR and used a sample of 38 clusters along 
with SNe Ia observations to test the CDDR. 
Assuming a parametrization for a possible deviation from the CDDR, $\eta = 1 + \eta_0 z/(1+z)$,  they found $\eta_0 = -0.07 \pm 0.24$ at 95.4\% C L. 

Recently, cluster mass data from  91 galaxy clusters detected via the Sunyaev-Zel'dovich effect (SZ; Sunyaev \& Zel'dovich 1970) at 148 GHz was reported by the Atacama Cosmology Telescope (ACT) survey (Hasselfield {\it{et al.}} 2013). It is important to emphasize that the gas mass fraction measurements of galaxy clusters depend on the model for the physics of the intracluster gas. The ACT team  adopted four models (see Sec. III for details) in order to estimate the corresponding cluster mass $M_{500}$, defined as the mass measured within the radius $R_{500}$, at which the enclosed  mean density is 500 times the critical density at the cluster redshift. In order to obtain $f_{gas}$ estimates (one of the main quantities for our analysis) for each cluster in the sample, we used a semi-empirical relation presented by Vikhlinin {\it{et al.}} (2009), where the observed gas fraction is a function of the total mass, $f_{gas} = f_{gas}(M)$. This relation was defined upon an observational sample where the total mass 
and the gas mass of the clusters were obtained through two different approaches. The first one is a direct measurement of the total mass and gas mass from observed ICM parameters (Kravtsov {\it{et al.}} 2006) whereas the second approach consists in measuring the total mass from the average temperature with the gas mass being determined from the X-ray image (Mohr {\it{et al.}} 1999).

Our goal in this paper is not only derive new bounds on a possible deviation from Eq. (1) using the current ACT and SNe Ia data but also to verify the robustness of the methods used to infer $M_{500}$. In other words, we will discuss the compatibility of these methods in light of the CDDR and the assumptions behind it. The paper is organized as follows. In Sec. 2 we describe the observational quantities used in this work.  The corresponding constraints on the departures of the  CDDR  are investigated in Sec. 3. We end this paper summarizing our main results in the conclusions section 4.

\section{Observational test}

We combine two observational data sets to perform our CDDR test, namely the gas mass fraction ($f_{gas}$), obtained from the $M_{500}$ data sets provided by 
the ACT galaxy clusters survey, and measurements of distance moduli from the Union2.1 SNe Ia compilation.

\subsection{$f_{gas}$}

The so-called gas mass fraction test (Sasaki 1996, Allen {\it{et al.}} 2002, 2004, 2008; Ettori {\it{et al.}} 2009) essentially assumes that the ratio $M_{gas}/M_{tot}$ is constant over the cosmic history, where $M_{gas}$ is the gas mass and $M_{tot}$ is the total mass (including dark matter) of the galaxy cluster. This assumption is quite reasonable, based on the fact there is no mechanism known able to push off the gas from the gravitational potential of the galaxy cluster. 

It is well known that the $f_{gas}$ obtained from X-ray measurements is given by (Sasaki, 1996)
\begin{equation}
f_{gas} \propto d_A^{3/2}\;,
\end{equation}
which is valid only when the CDDR is assumed. The general expression for the gas mass fraction is given by (Gon\c calves {\it{et al.}}, 2012)
\begin{equation}
\label{fgas} \label{fgas}
f_{gas}^{obs}(z) = n\frac{d_L^{*1/2}d_A^{*}}{d_L^{1/2}d_A} \; \; ,
\end{equation}
where the symbol $*$ denotes quantities that were obtained for the fiducial model assumed in the observations and the parameter $n$ defines the astrophysical modeling of the cluster. Many parameterizations are assumed in literature (Allen {\it{et al.}} 2002, Allen {\it{et al.}} 2008, Ettori {\it{et al.}} 2009, among others), but as we aim to constraint the CDD relation, $n$ is not an important quantity to our analysis and we marginalize over it. Note also that, along with the assumption of a constant $M_{gas}/M_{tot}$ ratio above mentioned, the distance ratio of Eq. (\ref{fgas}) accounts for deviations in the geometry of the universe from the fiducial model, which makes possible to use $f_{gas}$ data to test different cosmologies (Lima {\it{et al.}} 2003).

\subsubsection{f$_{\mbox{\rm gas}}$ from the ACT catalog}

As is widely known, properties of galaxy clusters encode information of the growth of structures in the universe. For this, data from optical and X-ray cluster surveys are currently used to constrain cosmological parameters. Recently, however, a new window of information regarding galaxy cluster physics was opened with 
the release of cluster surveys~(e.g., Staniszewski {\it{et al.}} 2009; Marriage {\it{et al.}} 2011; Williamson {\it{et al.}} 2011; Planck Collaboration 2011;  Reichardt {\it{et al.}} 2013) which make use of the SZ effect, a signature that does not diminish with luminosity distance because is nearly redshift-independent. In fact, revealed as a spectral distortion, thermal SZ signal is a tracer of the total thermal energy of the hot ($\sim 10^{8}$K) intracluster gas, therefore it is correlated to the mass of the cluster. 

\begin{figure*}
\center{\psfig{figure=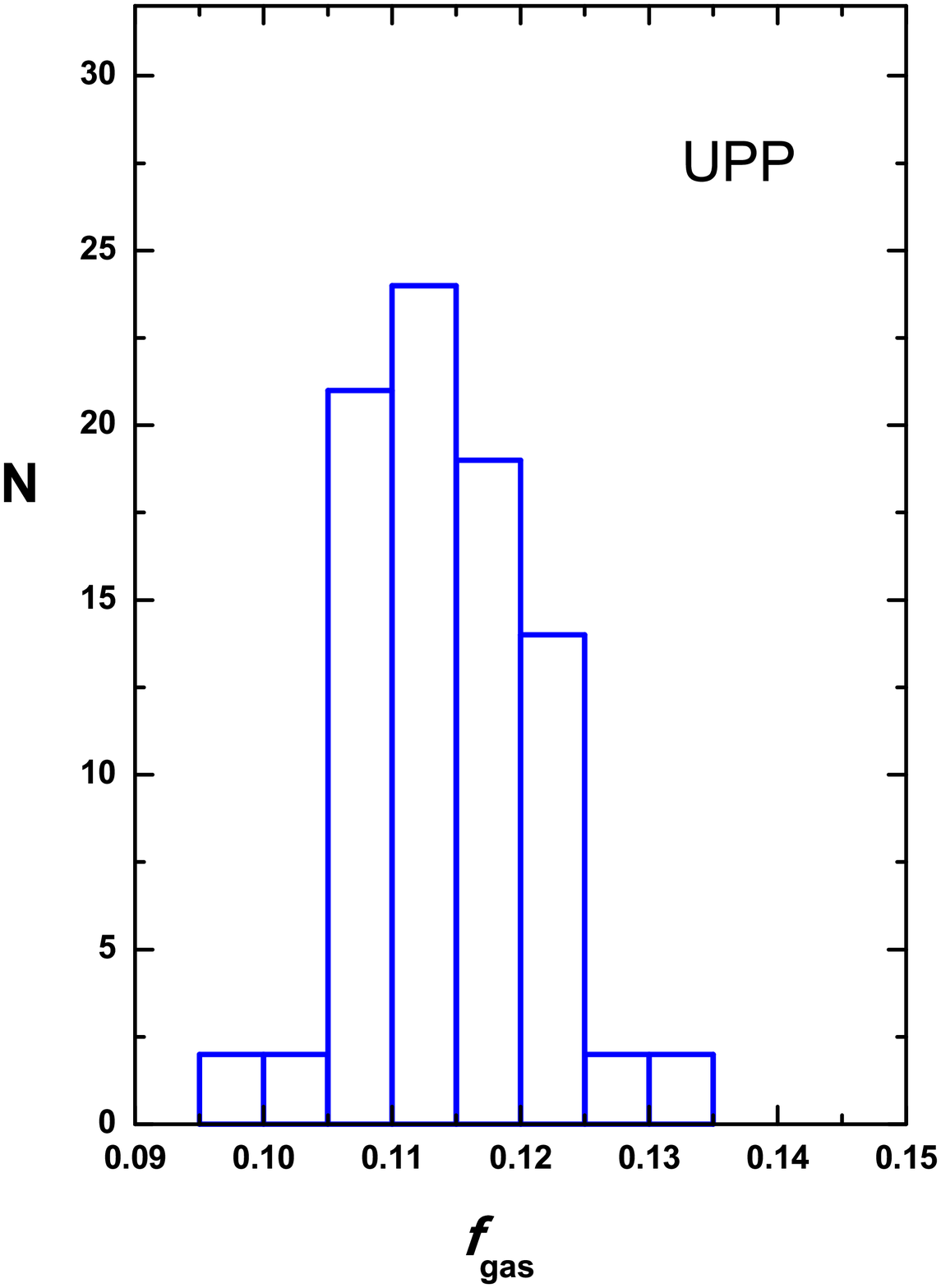,width=1.75in,height=2.1in}
\psfig{figure=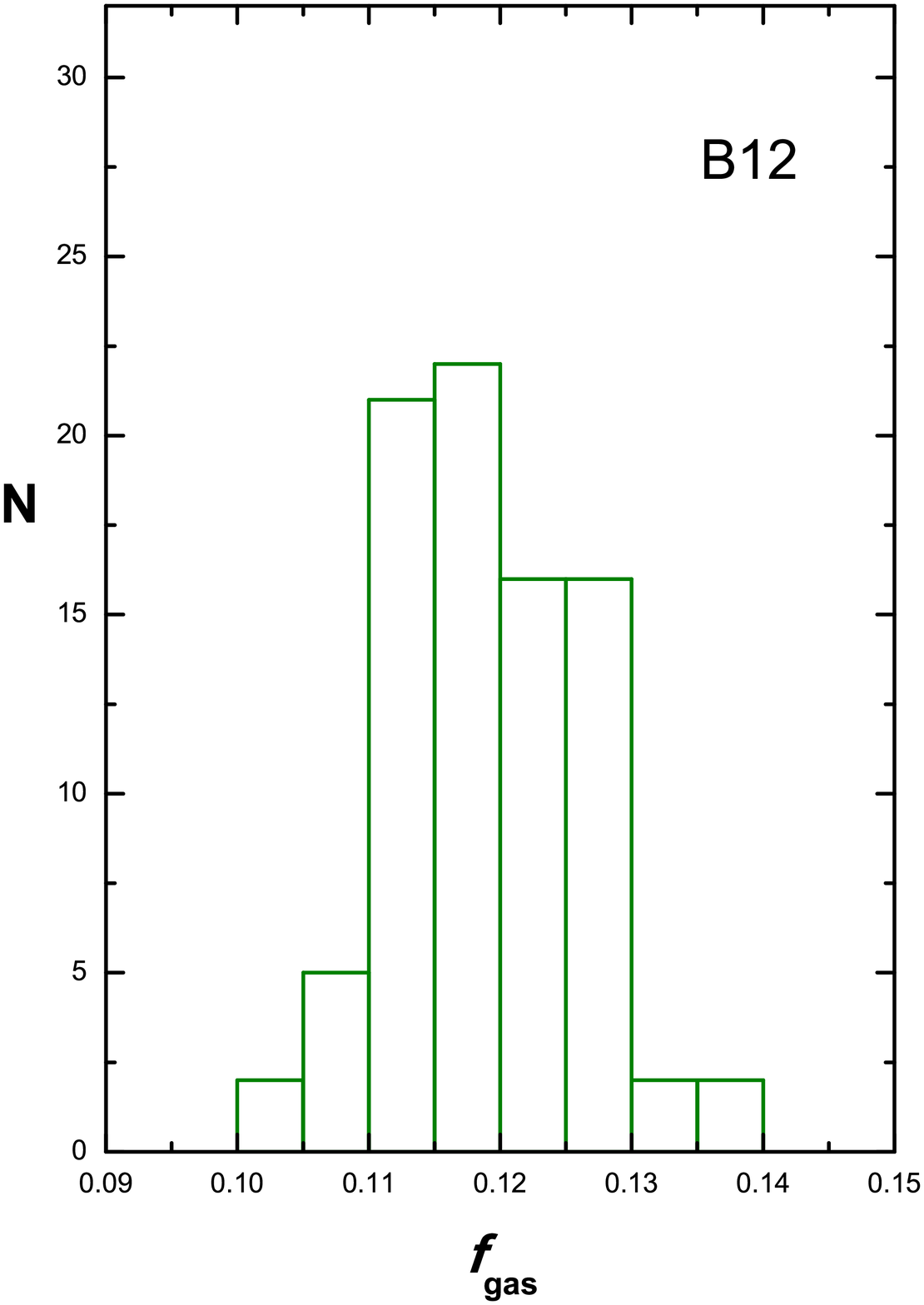,width=1.75in,height=2.1in}
\psfig{figure=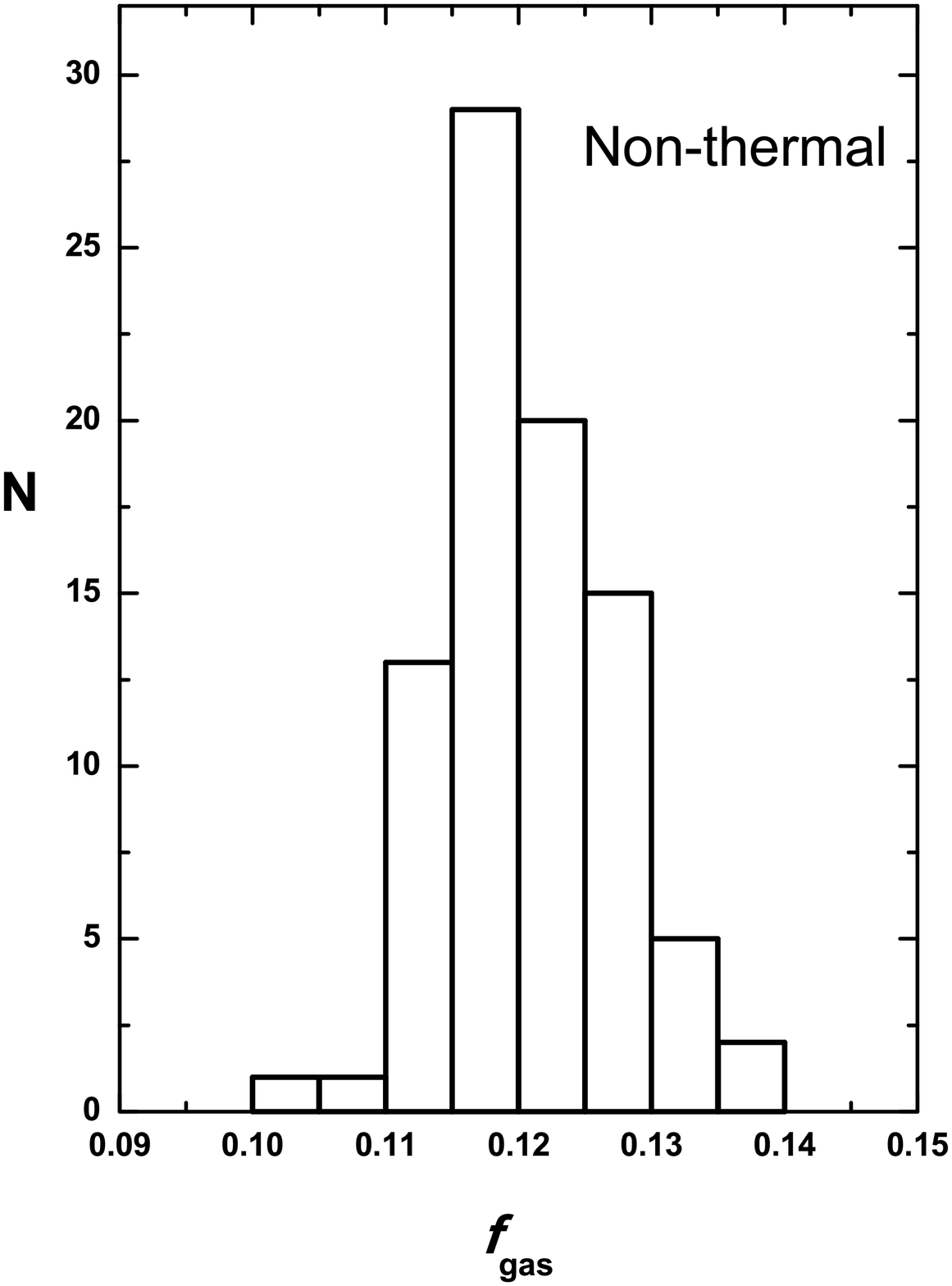,width=1.75in,height=2.1in}
\psfig{figure=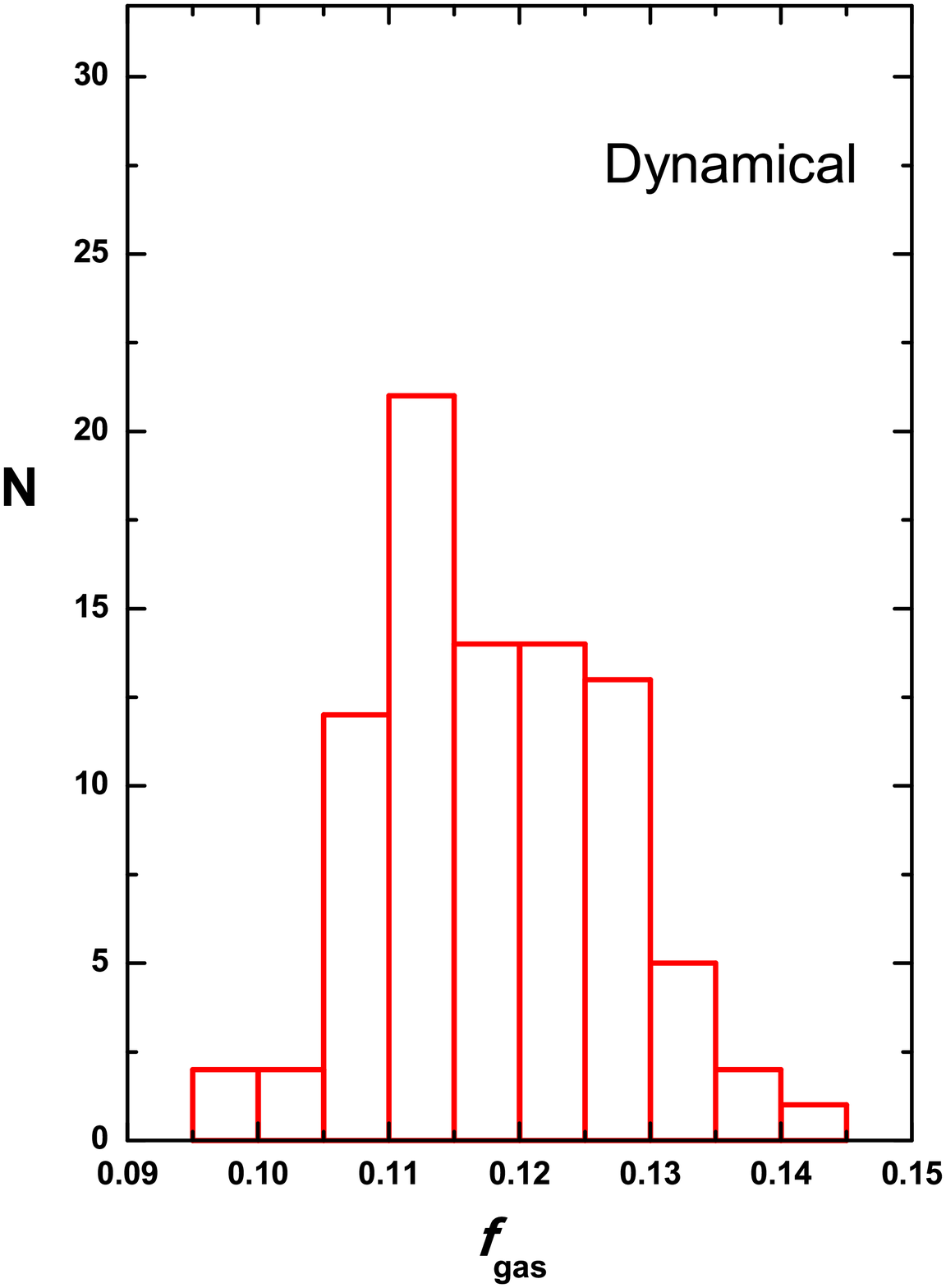,width=1.75in,height=2.1in}}
\caption{
The distribution of the number of clusters according to their gas mass fraction $f_{gas}$. 
The $f_{gas}$ values were inferred from the $M_{500}$ data (Hasselfield {\it{et al.}} 2013) according to the four methods discussed in the text. 
}
\end{figure*}

However, a model for the physics of the intracluster gas has to be assumed.  Recently, the ACT team investigated a new approach to obtain the cluster mass from cluster signal in filtered SZ maps. They adopt first a one-parameter family of {Universal Pressure Profiles} (UPP) as a baseline model for the intracluster gas pressure profile and apply it to their cluster sample, in order to estimate the corresponding cluster mass (Arnaud {\it{et al.}} 2010). This approach includes a mass dependence in the profile shape, which has been calibrated to X-ray observations using local clusters ($z < 0.2$). In this way, following this UPP approach, the cluster mass of 
the sample are measured within a characteristic radius with respect to the critical density such that, 
e.g., $M_{500}$ is defined as the mass measured within the radius, $R_{500}$, at which the enclosed 
mean density is 500 times the critical density at the cluster redshift. 
In this case, the clusters mass obtained using this method are termed 
$M_{500}^{\mbox{\footnotesize\sc upp}}$. 

As mentioned earlier, from the total mass it is possible to obtain $f_{gas}$ using a semi-empirical relation discussed by Vikhlinin {\it{et al.}} (2009). Such a relation can be written as
\begin{equation}
f_{gas} = 0.132 + 0.039\log M_{15} \;\; ,
\end{equation}
where $M_{15}$ is the cluster total mass $M_{500}$ in units of $10^{15} h^{-1} M_{\odot}$. 

Beside this UPP model procedure, other three approaches have been adopted by the ACT team to study the scaling relations that allow us to estimate $M_{500}$ from the cluster SZ signal strength. These are based on (i) structure formation simulations (Bode {\it{et al.}} 2012), where the density and temperature of the intracluster are modeled as a virialized ideal gas ($M_{500}^{\mbox{\tiny\rm B12}}$); (ii) using a non-thermal pressure and adiabatic model for the gas (Trac {\it{et al.}} 2011) ($M_{500}^{\mbox{\footnotesize\rm non-thermal}}$); and (iii) a dynamical estimate of the cluster mass using the galaxy velocity dispersions (Sifon {\it{et al.}} 2012) ($M_{500}^{\mbox{\footnotesize\rm dyn}}$). It is worth mentioning that, assuming the so-called concordance cosmology, a flat $\Lambda$CDM model with $\Omega_m = 0.3$, the scaling from UPP is nearly identical to the adiabatic non-thermal model, while a model incorporating non-thermal pressure is in 
better agreement with dynamical mass measurements (Hasselfield {\it{et al.}}, 2013). Considering these four methods to estimate $M_{500}$, in Fig. 1 we show the distribution of the number of clusters $N$ according to their gas mass fraction $f_{gas}$ inferred from the ACT sample. 

\subsection{SNe Ia}

In order to obtain measurements of $d_L$, we use the distance moduli ($\mu$) obtained from current SNe Ia observations. This quantity is related to the luminosity distance by
\begin{equation}
\label{mu}
\mu =  5 \log_{10}(\frac{d_L}{Mpc}) + 25\; .
\end{equation}
The data set used for SNe Ia is the Union 2.1 compilation (Suzuki {\it{et al.}} 2011) which contains 580 points distributed in the redshift range $0.01 < z < 1.41$. The SNe Ia redshifts were carefully chosen to coincide with the ones of the associated galaxy cluster sample with $\Delta z = |z_{cluster} - z_{SNe Ia}| \leq 0.01$ (Fig. 2) -- For more details on SNe Ia analysis we refer the reader to Suzuki {\it{et al.}} (2011).

\section{Analysis and Results}

In order to analyze the validity of the CDDR, we rewrite Eq. (1) as
$\frac{d_L}{d_A(1+z)^2} = \eta(z)$, 
where $\eta(z)$ quantifies a possible deviation from the CDDR. We assume in our analysis $\eta(z) = 1 + \eta_0 z$, which is compatible with the cosmographic limit $\eta(z \ll 1) = 1$. Combining the above equation with  (\ref{fgas}) and (\ref{mu}), we obtain (Gon\c calves {\it{et al.}} 2012)
\begin{equation}
\label{CDDRobs}
\eta_{obs} (z)=  \frac{ 10^{\frac{3}{5}(\mu - 25)} f_{gas}^2}{N^{2}(1+z)^6 d_A^{*3}}   \;,
\end{equation}
which is the observed quantity built from measurements of $\mu(z)$ and $f_{gas}$. Thus, the likelihood  estimator is determined by a $\chi^{2}$ statistics
$\chi^{2} = \sum_{i}\left[\eta(z_i) - \eta_{obs}(z_i)
\right]^{2}/{\sigma^2_{\eta_{obs}}}$,  
where $\sigma^2_{\eta_{obs}}$ takes into account the propagation of the statistical errors in Eq. (4) As mentioned earlier, the normalization factor $N$ [see Eq. (2)] is taken as a nuisance parameter so that we marginalize over it. 

\begin{figure}
\center{\psfig{figure=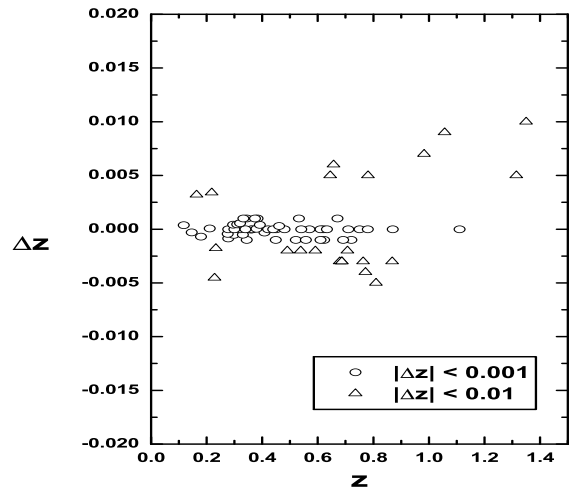,width=3.3in,height=3.0in}}
\caption{The redshift difference between SNe Ia and galaxy clusters used in the analysis.}
\end{figure}

The first results of this analysis are presented in Table 1. All methods considered are compatible with the CDDR at 2$\sigma$ level. For comparison, we also show the values of the reduced $\chi^2$ ($\chi^2/\nu$, where $\nu$ is the degree of freedom) and $p$-value. This latter refers to the probability of obtaining a test  at least as extreme as the observed one, under the assumption that the null hypothesis, i.e. $\eta_0 = 0$, is true.  For instance, $p < 0.05$ indicates that the null hypothesis is probably false (Jeffreys, 1961; Robert {\it{et al.}} 2009). Clearly, this is not the case for the results presented above -- for all $M_{500}$ methods considered we find values of $p$-value larger than 0.7.

In the analysis above, the maximum difference between the redshifts of the clusters and the SNIa  is $\Delta z \leq 0.01$. We also performed the same analysis restricting our data set to a subset where $\Delta z \leq 0.001$. Note that this is an important aspect for analyses of the CDDR involving different types of observables since the reciprocity theorem is valid for sources at the very same redshift. The number of points in this new subset decreases to 55 points whereas the $\chi^2$ improves about $15\%$. The mean values remain quite the same but the validity of CDDR can be verified even considering $1 \sigma$ error, as we can see in Table 2. 

An important aspect worth mentioning is that the likelihoods for $\eta_0$ (regardless of the method considered) have  values preferably negative, indicating that $\eta_0^{obs}$ may be underestimated and favoring a reduction of $d_L$ relative to $d_A$ [see Eq. (1)]. This is in full agreement with the results obtained by Holanda {\it et al.} (2012a) which uses only measurements of $f_{gas}$ from the SZ and X-ray observations. Physically, such result can be explained by a radiative process which increases the number of photons in a light bundle and therefore increases the apparent luminosity making the source appear closer (we refer the reader to Avgoustidis  2010, 2012 for more on possible coupling of exotic particles with photons).  Another explanation would be a possible excess of brightness of the SNe Ia data or a value not suitable for the galaxy clusters borders  ($r_{500}$) which may underestimate the gas mass fraction calculated in Eq. (4).


\begin{table}
\begin{center}
\begin{tabular}{lccc}
\hline
Data set 			& $\eta_0$ 			& $\chi^2/\nu$	& p-value \\
\hline \\
$f_{gas}^{UPP}$	+ SNIa		& $-0.08^{+0.11}_{-0.10}$	&	$0.74$ & 0.9645  \\
$f_{gas}^{B12}$	+ SNIa		& $-0.12^{+0.12}_{-0.09}$	&	$0.75$ & 0.9576 \\
$f_{gas}^{Non}$	+ SNIa		& $-0.17^{+0.11}_{-0.08}$	&	$0.75$ & 0.9576 \\
$f_{gas}^{Dyn}$	+ SNIa		& $-0.13^{+0.11}_{-0.10}$	&	$0.78$ & 0.9307 \\
\hline
\end{tabular}
\caption{Bounds on $\eta_0$ from $f_{gas}$ obtained using four different methods, as explained in the text. The SNe Ia data used in the analysis correspond to the Union2.1 compilation. The maximum difference between $z_{cluster}$ and $z_{SNe Ia}$ is $\Delta z \leq 0.01$. Error bars correspond to 1$\sigma$.}
\end{center}
\end{table}

\begin{table}
\begin{center}
\begin{tabular}{lccc}
\hline
Data set 			& $\eta_0$ 			& $\chi^2/\nu$	                    & p-value \\
\hline \\
$f_{gas}^{UPP}$	+ SNIa		& $-0.11^{+0.19}_{-0.16}$	&	$0.86$ & 0.7286  \\
$f_{gas}^{B12}$	+ SNIa		& $-0.12^{+0.17}_{-0.17}$	&	$0.86$ & 0.7286  \\
$f_{gas}^{Non}$	+ SNIa		& $-0.17^{+0.16}_{-0.18}$	&	$0.87$ & 0.7092  \\
$f_{gas}^{Dyn}$	+ SNIa		& $-0.14^{+0.19}_{-0.17}$	&	$0.87$ & 0.7092  \\
\hline
\end{tabular}
\caption{The same as in Table 1 considering $\Delta z \leq 0.001$.}
\end{center}
\end{table}


\section{Conclusions}

In this paper, we have explored if the four physical models used by the Atacama Cosmology 
Telescope to describe the intracluster gas of 91 galaxy clusters are compatible with the validity of the so-called Cosmic Distance Duality Relation (CDDR), $d_L/d_A(1+z)^2 = 1$. For this purpose, we aliviate the equality to $\frac{d_L}{d_A(1+z)^2} = \eta$, where $\eta$ is a time-dependent parametrization.  In our analysis, we have adopted a linear parametrization written as $\eta(z) = 1 + \eta_0 z$ for which values apart from $\eta_0 = 0$ leads to a violation of CDDR. To obtain the observational values for $d_L$ we have used measurements of distance moduli from SNe Ia (Union 2.1) whereas for $d_A$ we have used 91 measurements of gas mass fraction from galaxy clusters following Gon\c{c}alves {\it et al.} (2012).  

The statistical analysis performed showed that the results are almost independent of the method used to obtain $M_{gas}$. The main result of this analysis was performed by restricting the difference between $z_{cluster}$ and $z_{SNe Ia}$ to $\Delta z \leq 0.001$. In this case, the validity of the CDDR can be probed within 1$\sigma$ level irrespective of the $M_{500}$ method adopted. These results reinforce the interest in probing the CDDR using new and more precise techniques since a clear departure from $\eta_0 = 0$ may be associated either with an exotic gravity theory or a non-conservation of the photon number along the cosmic history.

\acknowledgements

The authors thank CNPq, INCT-A, INEspa\c{c}o and FAPERJ for the grants under which this work was carried out.

\label{lastpage}

\begin{thebibliography}{99}

\bibitem{Allen02} Allen S. W., Schmidt R. W. \& Fabian A. C., 2002, MNRAS, 334, L1
\bibitem{Allen08} Allen S. W., et al., 
2008, MNRAS, 383, 879
\bibitem{Arnaud10} Arnaud, M., et al. 2010, A\&A, 517, A92 (A10)
\bibitem{Avgoustidis10} Avgoustidis, A., Burrage, C., Redondo, J., Verde, L. \& Jimenez, R. 2010, JCAP, 10, 024
\bibitem{Avgoustidis12} Avgoustidis, A., Luzzi, G., Martins, C.J.A.P. \& Monteiro, A.M.R.V.L. 2012, JCAP, 2, 013
\bibitem{Basset04} Basset, B. A., \& Kunz, M. 2004, Phys. Rev. D, 69, 101305
\bibitem{Bode12} Bode, P., Ostriker, J. P., Cen, R., \& Trac, H. 2012, ApJ submitted (arXiv:1204.1762) (B12)
\bibitem{Brax13} Brax, P., Burrage, C., Davis, A.-C. \& Gubitosi, G., 2013, JCAP, 11, 001 
\bibitem{Csaki02} Csaki, C., Kaloper, N. \& J. Terning, 2002, Phys. Rev. Lett. 88, 161302.
\bibitem{daly}Daly, R. A. \& Djorgovski, S.G., 2003, ApJ, 597, 9
\bibitem{Ellis13} Ellis, G.F.R., Poltis, R., Uzan, J.-P. Weltman, A., 2013, 87, 10
\bibitem{Ellis71} Ellis G. F. R. 1971, Relativistic Cosmology, Proc. Int. School Phys. Enrico Fermi, ed. R. K. Sachs (New York: Academic Press), 104; reprinted in Gen. Rel. Grav., 41, 581, 2009
\bibitem{Ellis07} Ellis, G. F. R. 2007, Gen. Rel. Grav., 39, 1047 
\bibitem{Etherington33} Etherington, I. M. H. 1933, Phil. Mag., 15, 761; reprinted in Gen. Rel. Grav., 39, 1055, 2007
\bibitem{tori}Ettori, S. {\it et al.}, 2009, A\&A, 501, 61
\bibitem{gur}Gurvitz, L. I., 1994, ApJ, 425, 442
\bibitem{gurv}Gurvitz, L. I., Kellermann, K. I., \& Frey, S., 1999, A\&A, 342, 378
\bibitem{gon}Gon\c{c}alves, R. S., Holanda, R. F. L., \&  Alcaniz, J. S., 2012, MNRAS. 420, L43
\bibitem{ACT1} Hasselfield, M., Hilton, M., Marriage, T. A., et al., 2013, JCAP, 07, 008 
\bibitem{Holanda10} Holanda, R.F.L., Lima, J.A.S. \& Ribeiro, M.B. 2010, ApJ, 722, L233
\bibitem{ha}Holanda, R. F. L., Gon\c{c}alves, R. S. \& Alcaniz, J. S., 2012a,  JCAP {1206}, 022. [arXiv:1201.2378]
\bibitem{h3a} Holanda, R. F. L., Lima, J. A. S. \& Ribeiro M. B., 2012b, A\&A, 538,  131
\bibitem{jeffreys} Jeffreys, H. Theory of probability, 3rd edn , Oxford Classics series (reprinted 1998) (Oxford University Press, Oxford, UK, 1961). 
\bibitem{khe}Khedekar, S.  \& Chakraborti, S., 2011, PRL, 106, 22
\bibitem{Kravtsov06} Kravtsov, A. V., Vikhlinin, A., \& Nagai, D. 2006, ApJ, 650, 128
\bibitem{li}Li, Z., Wu, P. \& Yu, H., 2011, 729, L14
\bibitem{Lima:2001zz} Lima, J. A. S. \& Alcaniz, J. S., 2002, ApJ 566, 15 
\bibitem{Lima:2003dd} Lima, J. A. S., Cunha, J. V. \& Alcaniz, J. S., 2003, Phys.\ Rev.\ D68, 023510
\bibitem{lima}Lima, J. A. S., Cunha, J. V. \& Zanchin, V. T., 2012,
ApJL, 742, L26
\bibitem{Marriage} Marriage, T. A., {\it{et al.}}, 2011, ApJ, 737, 61
\bibitem{Meng12} Meng, X.-L., Zhang, T.-J., \& Zhan, H., 2012, ApJ, 745, 98
\bibitem{Mohr99} Mohr, J. J., Mathiesen, B., \& Evrard, A. E., 1999, ApJ, 517, 627
\bibitem{Nair11} Nair, R., Jhingan, S., \& Jain, D., 2011, JCAP, 05, 023
\bibitem{nam}  Nam, L. {\it et al.}, 2013, MNRAS, 436, 1017
\bibitem{PLA2011} Planck Collaboration, 2011, A\&A, 536, A8
\bibitem{Reichardt} Reichardt, C. L., et al., 2013, ApJ, 763, 127
\bibitem{robert} Robert, C. P., Chopin, N \& Rosseau, J., 2009, Statistical Science, 24, 141
\bibitem{Sarazin11} Sarazin C. L., 1988, X-Ray Emission from Cluster of Galaxies, 
Cambridge Univ. Press, Cambridge
\bibitem{Sasaki96} Sasaki S., 1996, PASJ, 48, L119
\bibitem{Sifon12} Sif\'on, C., {\it{et al.}},2012, ApJ submitted (arXiv:1201.0991) (S12)
\bibitem{Staniszewski} Staniszewski, Z., {\it{et al.}},2009, ApJ, 701, 32
\bibitem{SZ1970} Sunyaev, R. A. and Zel'dovich, Y. B., 1970, 
Comments on Astrophysics and Space Physics, 2, 66 
\bibitem{Suzuki12} Suzuki, N., Rubin, D., Lidman, C., {\it{et al.}},2012, ApJ, 746, 85

\bibitem{Trac11} Trac, H., Bode, P., \& Ostriker, J. P., 2011, ApJ, 727, 94
\bibitem{uzam}Uzan, J.P., Aghanim, N. and Mellier, Y., 2004, PRD, 70, 083533

\bibitem{Vikhlinin09} Vikhlinin, A., et al., 2009, ApJ, 692, 1033–1059
\bibitem{Vikhlinin06} Vikhlinin, A., et al., 2006, ApJ, 640, 691

\bibitem{Williamson} Williamson, R., {\it{et al.}}, 2011, ApJ, 738, 139
\bibitem{xi}  Xi, Y., Hao-Ran, Y., Zhi-Song, Z. \&  Tong-Jie, Z. 2013, APJ, 777, L24
\end{thebibliography}
\end{document}